# Black-Box Complexity: Breaking the $O(n \log n)$ Barrier of LeadingOnes

Benjamin Doerr and Carola Winzen

Max Planck Institute for Informatics, Saarbrücken, Germany

**Abstract**

We show that the unrestricted black-box complexity of the $n$-dimensional XOR- and permutation-invariant LeadingOnes function class is $O(n \log(n)/ \log \log n)$. This shows that the recent natural looking $O(n \log n)$ bound is not tight.

The black-box optimization algorithm leading to this bound can be implemented in a way that only 3-ary unbiased variation operators are used. Hence our bound is also valid for the unbiased black-box complexity recently introduced by Lehre and Witt. The bound also remains valid if we impose the additional restriction that the black-box algorithm does not have access to the objective values but only to their relative order (ranking-based black-box complexity).

## 1 Introduction

The black-box complexity of a set $\mathcal{F}$ of functions $\mathcal{S} \to \mathbb{R}$, roughly speaking, is the number of function evaluations necessary to find the maximum of any member of $\mathcal{F}$ which—apart from the points evaluated so far—is unknown. This and related notions are used to describe how difficult a problem is to be solved via general-purpose (randomized) search heuristics. Consequently, black-box complexities are very general lower bounds which are valid for a wide range of evolutionary algorithms. A number of different black-box notions exist, each capturing different classes of randomized search heuristics, cf. [DJW06], [LW10], and [DW11].

In this paper we consider the unrestricted black-box model by Droste, Jansen, and Wegener [DJW06] and the unbiased black-box model by Lehre and Witt [LW10], and we shortly remark on the ranking-based models which we propose in [DW11]. A formal definition of the first two models is given in Section 2. For now, let us just mention that algorithms in the unrestricted black-box model are allowed to query any bit string, whereas in the $k$-ary unbiased model, an algorithm may only query search points sampled from a $k$-ary unbiased distribution. That is, in each iteration, the algorithm may either query a random search point or, based upon at most $k$ previously queried search points, it may generate a new one. The new search point can be generated only by using so-called unbiased variation operators. These are operators that are symmetric both in the bit values (0 or 1) and the bit positions. More formally, the variation operation must be invariant under all automorphisms of the hypercube.

In this work, we are concerned with the black-box complexity of the LeadingOnes function, which is one of the classical test functions for analyzing the optimization behavior of different search heuristics. The function itself is defined via $\text{Lo} : \{0,1\}^n \to [0..n], x \mapsto \max\{i \in [0..n] \mid \forall j \leq i : x_j = 1\}$. It was introduced in [Rud97] to disprove a previous conjecture by Mühlenbein [Müh92] that any unimodal function can be optimized by the well-known (1+1) evolutionary algorithm (EA) in $O(n \log n)$ iterations. Rudolph [Rud97] proves an upper bound of $O(n^2)$ for the expected optimization time of the $(1 + 1)$ EA on LO and concludes from experimental studies a lower bound of $\Omega(n^2)$—a bound which was rigorously proven in 2002 by Droste, Jansen, and Wegener [DJW02]. This $\Theta(n^2)$ expected optimization time of the simple $(1 + 1)$ EA seems optimal among the commonly studied evolutionary algorithms.

Note that the unrestricted black-box complexity of the LO function is 1: The algorithm querying the all-ones vector $(1, \ldots, 1)$ in the first query is optimal. Motivated by this and by the fact that the unbiased black-box model only allows variation operators which are invariant with respect to the bit values and the bit positions, we shall study here a generalized version of the LO function. More precisely, we consider the closure of LO under all permutations $\sigma \in S_n$ and under all exchanges of the bit values 0 and 1. It is immediate, that each of these functions has a fitness landscape that is isomorphic to the one induced by LO. To be more precise, we define for any bit string $z \in \{0,1\}^n$ and any permutation $\sigma$ of $[n]$ the function

$$\text{Lo}_{z,\sigma} : \{0,1\}^n \to [0..n], x \mapsto \max\{i \in [0..n] \mid \forall j \leq i : z_{\sigma(j)} = x_{\sigma(j)}\}.$$

We let LEADINGONES$_n$ be the set $\{\text{Lo}_{z,\sigma} \mid z \in \{0,1\}^n, \sigma \in S_n\}$ of all such functions.

Note that this definition differs from the one in [DJW06], where only the subclass LEADINGONES$_n^0 := \{\text{Lo}_{z,\text{id}} \mid z \in \{0,1\}^n\}$ not invariant under permutations of $[n]$ is studied. Here, id denotes the identity mapping on $[n]$. For this restricted subclass, Droste, Jansen, and Wegener [DJW06] prove an unrestricted black-box complexity of $\Theta(n)$. Of course, their lower bound $\Omega(n)$ is a lower bound for the unrestricted black-box complexity of the general LEADINGONES$_n$ function class, and consequently, a lower bound for the unbiased black-box complexity of LEADINGONES$_n$.

The function class LEADINGONES$_n$ has implicitly been studied for the first time in [LW10], where Lehre and Witt show that indeed the (1+1) EA is provably (asymptotically) optimal among all unbiased black-box algorithms of arity at most one. This establishes a natural $\Theta(n^2)$ bound for LEADINGONES$_n$.

Surprisingly, it turns out that this bound does not hold for the unrestricted black-box model and that it does not even hold in the 2-ary unbiased black-box model. In [DJK$^+$11] it is shown that, assuming knowledge on $\sigma(1), \ldots, \sigma(\ell)$, one can perform a binary search to determine $\sigma(\ell + 1)$ and its corresponding bit value. Since this has to be done $n$ times, an upper bound of $O(n \log n)$ for the unrestricted black-box complexity of LEADINGONES$_n$ follows. Furthermore, this $O(n \log n)$ bound can already be achieved in the binary unbiased model. Up to now, this is the best known upper bound for the unrestricted and the 2-ary unbiased black-box complexity of LEADINGONES$_n$.

In this work we show that both in the unrestricted model (Section 3) and for arities at least three (Section 4), one can do better, namely that $O(n \log(n)/ \log \log n)$ queries suffice to optimize any function in LEADINGONES$_n$. This breaks the previous $O(n \log n)$ barrier. This result also shows why previous attempts to prove an $\Omega(n \log n)$ lower bound must fail.

Unfortunately, also the ranking-based model does not help to overcome this unnatural low black-box complexity. We shall comment in Section 5 that the 3-ary unbiased ranking-based black-box complexity of LEADINGONES$_n$, too, is $O(n \log(n)/ \log \log n)$.

As for the memory-restricted model we note without proof that a memory of size $O(\sqrt{\log n})$ suffices to achieve the same bound.

---
**Algorithm 1:** Scheme of an unrestricted black-box algorithm for optimizing $f : \mathcal{S} \to \mathbb{R}$
---
**1 Initialization:**
**2**     Sample $x^{(0)}$ according to some probability distribution $p^{(0)}$ on $\mathcal{S}$;
**3**     Query $f(x^{(0)})$;
**4 Optimization:**
**5**     **for** $t = 1, 2, 3, \ldots$ **do**
**6**        Depending on $\left((x^{(0)}, f(x^{(0)})), \ldots, (x^{(t-1)}, f(x^{(t-1)}))\right)$ choose a probability distribution $p^{(t)}$ on $\mathcal{S}$;
**7**        Sample $x^{(t)}$ according to $p^{(t)}$;
**8**        Query $f(x^{(t)})$;
---

## 2 Preliminaries

In this section we briefly introduce the two black-box models considered in this work, the *unrestricted black-box model* by Droste, Jansen, and Wegener [DJW06] and the *unbiased black-box model* by Lehre and Witt [LW10]. Due to space limitations, we keep the presentation as concise as possible. For a more detailed discussion of the two different black-box models and for the definition of the ranking-based versions considered in Section 5, we refer the reader to [DW11].

Before we introduce the two black-box models, let us fix some notation. For all positive integers $k \in \mathbb{N}$ we abbreviate $[k] := \{1, \ldots, k\}$ and $[0..k] := [k] \cup \{0\}$. By $e_k^n$ we denote the $k$-th unit vector $(0, \ldots, 0, 1, 0, \ldots, 0)$ of length $n$. For a set $I \subseteq [n]$ we abbreviate $e_I^n := \sum_{i \in I} e_i^n = \oplus_{i \in I} e_i^n$, where $\oplus$ denotes the bitwise exclusive-or. By $S_n$ we denote the set of all permutations of $[n]$ and for $x = (x_1, \ldots, x_n) \in \{0,1\}^n$ and $\sigma \in S_n$ we abbreviate $\sigma(x) := (x_{\sigma(1)}, \ldots, x_{\sigma(n)})$. For any two strings $x, y \in \{0,1\}^n$ let $\mathcal{B}(x, y) := \{i \in [n] \mid x_i = y_i\}$, the set of positions in which $x$ and $y$ coincide.

For $r \in \mathbb{R}_{\geq 0}$, let $\lceil r \rceil := \min\{n \in \mathbb{N}_0 \mid n \geq r\}$ and $\lfloor r \rfloor := \max\{n \in \mathbb{N}_0 \mid n \leq r\}$. For the purpose of readability we sometime omit the $\lceil \cdot \rceil$ signs, that is, whenever we write $r$ where an integer is required, we implicitly mean $\lceil r \rceil$. All logarithms $\log$ in this work are base two logarithms. By $\ln$ we denote the logarithm to the base $e := \exp(1)$.

**Black-Box Complexity.** Let $\mathcal{A}$ be a class of algorithms and $\mathcal{F}$ be a class of functions. For every $A \in \mathcal{A}$ and $f \in \mathcal{F}$ let $T(A, f) \in \mathbb{R} \cup \{\infty\}$ be the expected number of fitness evaluations until $A$ queries for the first time some $x \in \arg\max f$. We call $T(A, f)$ the *expected optimization time of $A$ for $f$*. The *$A$-black-box complexity of $\mathcal{F}$* is $T(A, \mathcal{F}) := \sup_{f \in \mathcal{F}} T(A, f)$, the worst-case expected optimization time of $A$ on $\mathcal{F}$. The *$\mathcal{A}$-black-box complexity of $\mathcal{F}$* is $\inf_{A \in \mathcal{A}} T(A, \mathcal{F})$, the best worst-case expected optimization time an algorithm of $\mathcal{A}$ can exhibit on $\mathcal{F}$.

**The Unrestricted Black-Box Model.** The black-box complexity of a class of functions depends crucially on the class of algorithms under consideration. If the class $\mathcal{A}$ contains all (deterministic and randomized) algorithms, we refer to the respective complexity as the *unrestricted black-box complexity*. This is the model by Droste, Jansen, and Wegener [DJW06]. The scheme of an unrestricted algorithm is presented in Algorithm 1.

Note that this algorithm runs forever. Since our performance measure is the expected number of iterations needed until for the first time an optimal search point is queried, we do not specify a termination criterion for black-box algorithms here.

**The Unbiased Black-Box Model.** As observed already by Droste, Jansen, and Wegener [DJW06], the unrestricted black-box complexity can be surprisingly low for different function classes. For example, it is shown in [DJW06] that the unrestricted black-box complex-ity of the NP-hard optimization problem MAXCLIQUE is polynomial.

**Algorithm 2:** Scheme of a $k$-ary unbiased black-box algorithm
1 **Initialization:**
2    Sample $x^{(0)} \in \{0,1\}^n$ uniformly at random;
3    Query $f(x^{(0)})$;
4 **Optimization:**
5    for $t = 1, 2, 3, \ldots$ do
6       Depending on $\left(f(x^{(0)}), \ldots, f(x^{(t-1)})\right)$ choose up to $k$ indices $i_1, \ldots, i_k \in [0..t-1]$ and a $k$-ary unbiased distribution $D(\cdot \mid x^{(i_1)}, \ldots, x^{(i_k)})$;
7       Sample $x^{(t)}$ according to $D(\cdot \mid x^{(i_1)}, \ldots, x^{(i_k)})$;
8       Query $f(x^{(t)})$;

This motivated Lehre and Witt [LW10] to define a more restrictive class of algorithms, the so-called *unbiased black-box model,* where algorithms may generate new solution candidates only from random or previously generated search points and only by using *unbiased* variation operators, cf. Definition 1. Still the model captures most of the commonly studied search heuristics, such as many $(\mu + \lambda)$ and $(\mu, \lambda)$ evolutionary algorithms, simulated annealing algorithms, the Metropolis algorithm, and the Randomized Local Search algorithm (confer the book [AD11] for the definitions of these algorithms).

**Definition 1** (Unbiased Variation Operator). *For all $k \in \mathbb{N}$, a $k$-ary unbiased distribution $\left(D(\cdot \mid y^{(1)}, \ldots, y^{(k)})\right)_{y^{(1)}, \ldots, y^{(k)} \in \{0,1\}^n}$ is a family of probability distributions over $\{0,1\}^n$ such that for all inputs $y^{(1)}, \ldots, y^{(k)} \in \{0,1\}^n$ the following two conditions hold.*

$$(i) \forall x, z \in \{0,1\}^n : D(x \mid y^{(1)}, \ldots, y^{(k)}) = D(x \oplus z \mid y^{(1)} \oplus z, \ldots, y^{(k)} \oplus z);$$
$$(ii) \forall x \in \{0,1\}^n \forall \sigma \in S_n : D(x \mid y^{(1)}, \ldots, y^{(k)}) = D(\sigma(x) \mid \sigma(y^{(1)}), \ldots, \sigma(y^{(k)})).$$

*We refer to the first condition as $\oplus$-invariance and to the second as permutation invariance. An operator sampling from a $k$-ary unbiased distribution is called a $k$-ary unbiased variation operator.*

1-ary—also called *unary*—operators are sometimes referred to as mutation operators, in particular in the field of evolutionary computation. 2-ary—also called *binary*—operators are often referred to as crossover operators. If we allow arbitrary arity, we call the corresponding model the $\ast$-ary unbiased black-box model.

A $k$-ary unbiased black-box algorithm can now be described via the scheme of Algorithm 2. The $k$-ary unbiased black-box complexity of some class of functions $\mathcal{F}$ is the complexity of $\mathcal{F}$ with respect to all $k$-ary unbiased black-box algorithms.

## 3   On LeadingOnes$_n$ in the Unrestricted Model

This section is devoted to the main contribution of this work, Theorem 1. Recall from the introduction that we have defined

$$\text{Lo}_{z,\sigma} : \{0,1\}^n \to [0..n], x \mapsto \max\{i \in [0..n] \mid \forall j \leq i : z_{\sigma(j)} = x_{\sigma(j)}\}$$

and LEADINGONES$_n := \{\text{Lo}_{z,\sigma} \mid z \in \{0,1\}^n, \sigma \in S_n\}$.

**Theorem 1.** *The unrestricted black-box complexity of* LEADINGONES$_n$ *is $O(n \log(n) / \log \log n)$.*

The proof of Theorem 1 is technical. For this reason, we split it into several lemmata. The main proof can be found at the end of this section. We remark already here that the algorithm certifying Theorem 1 will make use of unbiased variation operators only. Hence, it also proves that the $*$-ary unbiased black-box complexity of LeadingOnes$_n$ is $O(n \log(n)/\log \log n)$. This will be improved in Section 4.

The main idea of both the $*$-ary and the 3-ary algorithm is the following. Given a bit string $x$ of fitness $\text{Lo}_{z,\sigma}(x) = \ell$, we iteratively first learn $k := \sqrt{\log n}$ bit positions $\sigma(\ell+1), \ldots, \sigma(\ell+k)$ and their corresponding bit values which we fix for all further iterations of the algorithm. Learning such a block of size $k$ will require $O(k^3/\log k^2)$ queries. Since we have to optimize $n/k$ such blocks, the overall expected optimization is $O(nk^2/\log k^2) = O(n \log(n)/\log \log n)$. In what follows, we shall formalize this idea.

**Convention:** For all following statements let us fix a positive integer $n$, a bit string $z \in \{0,1\}^n$ and a permutation $\sigma \in S_n$.

**Definition 2** (Encoding Pairs). *Let $\ell \in [0..n]$ and let $y \in \{0,1\}^n$ with $\text{Lo}_{z,\sigma}(y) = \ell$. If $x \in \{0,1\}^n$ satisfies $\text{Lo}_{z,\sigma}(x) \geq \text{Lo}_{z,\sigma}(y)$ and $\ell = |\{i \in [n] \mid x_i = y_i\}|$, we call $(x,y)$ an $\ell$-encoding pair for $\text{Lo}_{z,\sigma}$.*

*If $(x,y)$ is an $\ell$-encoding pair for $\text{Lo}_{z,\sigma}$, the bit positions $\mathcal{B}(x,y)$ are called the $\ell$-encoding bit positions of $\text{Lo}_{z,\sigma}$ and the bit positions $j \in [n]\backslash\mathcal{B}(x,y)$ are called* non-encoding.

If $(x,y)$ is an $\ell$-encoding pair for $\text{Lo}_{z,\sigma}$ we clearly either have have $\text{Lo}_{z,\sigma}(x) > \ell$ or $\text{Lo}_{z,\sigma}(x) = \text{Lo}_{z,\sigma}(y) = n$. For each non-optimal $y \in \{0,1\}^n$ we call the unique bit position which needs to be flipped in $y$ in order to increase the objective value of $y$ the $\ell$-*critical bit position of* $\text{Lo}_{z,\sigma}$. Clearly, the $\ell$-critical bit position of $\text{Lo}_{z,\sigma}$ equals $\sigma(\ell+1)$, but since $\sigma$ is unknown to the algorithm we shall make use of this knowledge only in the analysis, and not in the definition of our algorithms. In the same spirit, we call the $k$ bit positions $\sigma(\ell+1), \ldots, \sigma(\ell+k)$ the $k$ $\ell$-*critical bit positions of* $\text{Lo}_{z,\sigma}$.

**Lemma 2.** *Let $\ell \in [0..n-1]$ and let $(x,y) \in \{0,1\}^n \times \{0,1\}^n$ be an $\ell$-encoding pair for $\text{Lo}_{z,\sigma}$. Furthermore, let $k \in [n - \text{Lo}_{z,\sigma}(y)]$ and let $y' \in \{0,1\}^n$ with $\ell \leq \text{Lo}_{z,\sigma}(y') < \ell + k$.*

*If we create $y''$ from $y'$ by flipping each non-encoding bit position $j \in [n]\backslash\mathcal{B}(x,y)$ with probability $1/k$, then*
$$\Pr[\text{Lo}_{z,\sigma}(y'') > \text{Lo}_{z,\sigma}(y')] \geq (ek)^{-1}.$$

*Proof.* First note that due to $\text{Lo}_{z,\sigma}(y') \geq \text{Lo}_{z,\sigma}(y)$, we clearly have $y'_j = x_j$ for all $\ell$-encoding bit positions $j \in \mathcal{B}(x,y)$. Since we do not allow these bit positions $\mathcal{B}(x,y)$ to be changed, we necessarily also have $\text{Lo}_{z,\sigma}(y'') \geq \ell$.

Let $c \in [0..\sqrt{\log n}]$ such that $\text{Lo}_{z,\sigma}(y') = \ell + c$. Then $\text{Lo}_{z,\sigma}(y'') > \text{Lo}_{z,\sigma}(y')$ if and only if both

(i) none of the $c$ bit positions $\sigma(\text{Lo}_{z,\sigma}(y)+1), \ldots, \sigma(\text{Lo}_{z,\sigma}(y)+c)$ is flipped, and

(ii) the $\ell + c$-critical bit position $\sigma(\ell+c+1)$ is flipped.

This yields
$$\Pr[\text{Lo}_{z,\sigma}(y'') > \text{Lo}_{z,\sigma}(y')] = (1-\tfrac{1}{k})^c \tfrac{1}{k} \geq (1-\tfrac{1}{k})^{k-1} \tfrac{1}{k} \geq (ek)^{-1},$$

where we make use of the well-known fact that for all positive integers $k$ it holds that $(1-\tfrac{1}{k})^{k-1} \geq e^{-1}$. □

**Algorithm 3:** The $(x,y)$-encoded $(1+1)$ evolutionary algorithm with mutation probability $1/k$.

**1 Input:** $\ell$-encoding pair $(x,y) \in \{0,1\}^n \times \{0,1\}^n$.
**2** $y' \leftarrow y$;
**3 while** $\mathrm{Lo}_{z,\sigma}(y') < \mathrm{Lo}_{z,\sigma}(y) + k$ **do**
**4** $\quad y'' \leftarrow \mathtt{random}(y', x, y, 1/k)$;
**5** $\quad$ Query $\mathrm{Lo}_{z,\sigma}(y'')$;
**6** $\quad$ **if** $\mathrm{Lo}_{z,\sigma}(y'') > \mathrm{Lo}_{z,\sigma}(y')$ **then** $y' \leftarrow y''$;
**7 Output** $y'$;

Lemma 2 motivates us to formulate Algorithm 3 which can be seen as a variant of the standard $(1+1)$ EA, in which we fix some bit positions and where we apply a non-standard mutation probability. The variation operator $\mathtt{random}(y', x, y, 1/k)$ samples a bit string $y''$ from $y'$ by flipping each non-encoding bit position $j \in [n]\setminus\mathcal{B}(x,y)$ with probability $1/k$. This is easily seen to be an unbiased variation operator of arity 3.

The following statement follows easily from Lemma 2 and the linearity of expectation.

**Corollary 3.** *Let $(x,y)$, $\ell$, and $k$ be as in Lemma 2. Then the $(x,y)$-encoded $(1+1)$ EA with mutation probability $1/k$, after an expected number of $O(k^2)$ queries, outputs a bit string $y' \in \{0,1\}^n$ with $\mathrm{Lo}_{z,\sigma}(y') \geq \ell + k$.*

A second key argument in the proof of Theorem 1 is the following. Given an $\ell$-encoding pair $(x,y)$ and a bit string $y'$ with $\mathrm{Lo}_{z,\sigma}(y') \geq \ell + \sqrt{\log n}$, we are able to learn the $\sqrt{\log n}$ $\ell$-critical bit positions $\sigma(\ell+1), \ldots, \sigma(\ell+\sqrt{\log n})$ in an expected number of $O(\log^{3/2}(n)/\log\log n)$ queries. This will be formalized in the following statements.

**Lemma 4.** *Let $\ell \in [0..n-\lceil\sqrt{\log n}\rceil]$ and let $(x,y)$ be an $\ell$-encoding pair for $\mathrm{Lo}_{z,\sigma}$. Furthermore, let $y'$ be a bit string with $\mathrm{Lo}_{z,\sigma}(y') \geq \ell + \sqrt{\log n}$.*

*For each $i \in [8e\log^{3/2}(n)/\log\log n]$ let $y^i$ be sampled from $y'$ by independently flipping each non-encoding bit position $j \in [n]\setminus\mathcal{B}(x,y)$ with probability $1/\sqrt{\log n}$.*

*For each $c \in [\sqrt{\log n}]$ let*

$$X_c := \{y^i \mid i \in [8e\log^{3/2}(n)/\log\log n] \text{ and } \mathrm{Lo}_{z,\sigma}(y^i) = \ell + c - 1\},$$

*the set of all samples $y^i$ with $\mathrm{Lo}_{z,\sigma}(y^i) = \ell + c - 1$.*

*Then*

$$\Pr\left[\forall c \in [\sqrt{\log n}] : |X_c| \geq 4\log(n)/\log\log n\right] \geq 1 - o(1).$$

*Proof.* For readability purposes, let us abbreviate $k := \sqrt{\log n}$.

First, let us consider the size $|X_c|$ for some fixed $c \in [k]$. By the same arguments as used in the proof of Lemma 2 we have for any $i \in [8e\log^{3/2}(n)/\log\log n]$ that

$$\Pr[\mathrm{Lo}_{z,\sigma}(y^i) = \ell + c - 1] = (1 - \tfrac{1}{k})^{c-1}\tfrac{1}{k} \geq (1 - \tfrac{1}{k})^{k-1}\tfrac{1}{k} \geq \tfrac{1}{ek}.$$

Thus, $\mathrm{E}[|X_c|] \geq 8\log(n)/\log\log n$ by the linearity of expectation.

If we set $Y_{i,c} = 1$ if $\mathrm{Lo}_{z,\sigma}(y^i) = \ell + c - 1$ and $Y_{i,c} = 0$ otherwise, we have $|X_c| := \sum_{i=1}^{8e\log^{3/2}(n)/\log\log n} Y_{i,c}$. In particular, $|X_c|$ is the sum of independent random variables. We can

thus apply Chernoff's bounds (cf. [AD11] for a compact introduction) to bound the deviation of $|X_c|$ from its expectation and obtain

$$\Pr\left[|X_c| < \tfrac{1}{2}\operatorname{E}[|X_c|]\right] \leq \exp\left(-\tfrac{1}{8}\operatorname{E}[|X_c|]\right) \leq \exp\left(-\log(n)/\log\log n\right) \leq 1/\log n\,,$$

where the last inequality follows from $\log(n)/\log\log n > \ln\log n$ for large enough $n$. By a simple union bound we conclude

$$\Pr\left[\forall c \in [\sqrt{\log n}] : |X_c| \geq 4\log(n)/\log\log n\right] \geq 1 - 1/\sqrt{\log n} = 1 - o(1)\,.$$

$\square$

These sets $X_c$ are large enough to identify $\sigma(\ell + c)$.

**Lemma 5.** *Let $\ell$, $(x, y)$, and $y'$ be as in Lemma 4 and let $t := 4\log(n)/\log\log n$.*

*For any $c \in [\sqrt{\log n}]$ let $X_c$ be a set of at least $t$ bit strings $y^1(c), \ldots, y^{|X_c|}(c)$ with fitness $\operatorname{Lo}_{z,\sigma}(y^i(c)) = \ell + c - 1$, which are sampled from $y'$ by independently flipping each non-encoding bit position $j \in [n] \setminus \mathcal{B}(x, y)$ with probability $1/\sqrt{\log n}$.*

*Then we have, with probability at least $1 - o(1)$, that for all $c \in [\sqrt{\log n}]$ there exists only one non-encoding $j := j_{\ell+c} \in [n] \setminus \mathcal{B}(x, y)$ with $y'_j = 1 - y^i_j(c)$ for all $i \in [|X_c|]$. Clearly, $j = \sigma(\ell + c)$.*

*Proof.* Let us first consider some fix value $c \in [\sqrt{\log n}]$. For any $i \in [|X_c|]$ we have, by definition, that $\operatorname{Lo}_{z,\sigma}(y^i(c)) = \ell + c - 1$. Thus, it must hold that $y^i_{\sigma(\ell+c)}(c) = 1 - y'_{\sigma(\ell+c)}$ and $y^i_{\sigma(j)}(c) = y'_{\sigma(j)}$ for all $j < \ell + c$. Let

$$\mathcal{I}_{\ell+c} := [n] \setminus (\mathcal{B}(x, y) \cup \{\sigma(\ell+1), \ldots, \sigma(\ell+c)\})$$
$$= [n] \setminus \{\sigma(1), \ldots, \sigma(\ell+c)\}\,.$$

Since all bit flips are mutually independent, we have for any $i \in [|X_c|]$ and any fixed $j \in \mathcal{I}_{\ell+c}$ that the entry $y^i_j(c)$ in the $j$-th bit position equals $1 - y'_j$ with probability $1/\sqrt{\log n}$. Note that this remains true despite the fact that we condition on $\operatorname{Lo}_{z,\sigma}(y^i) = \ell + c - 1$.

Thus, for any $j \in \mathcal{I}_{\ell+c}$ we have

$$\Pr[\forall i \in [|X_c|] : y^i_j(c) = 1 - y'_j] \leq (1/\sqrt{\log n})^{|X_c|}$$
$$\leq (1/\sqrt{\log n})^{4\log(n)/\log\log n}$$
$$= 2^{-2\log n} = 1/n^2\,.$$

By the union bound, the probability that there exists a $j \in \mathcal{I}_{\ell+c}$ with $y^i_j(c) = 1 - y'_j$ for all $i \in [|X_c|]$ is bounded from above by $1/n$.

And by again applying a union bound we find that

$$\Pr[\exists c \in [\sqrt{\log n}] \exists j \in \mathcal{I}_{\ell+c} \forall i \in [|X_c|] : y^i_j(c) = 1 - y'_j] \leq \sqrt{\log n}/n = o(1)\,.$$

$\square$

Combining Lemma 4 with Lemma 5 we immediately gain the following.

**Corollary 6.** *Let $\ell, (x, y), y'$, and $y^i, i = 1, \ldots, 8e\log^{3/2}(n)/\log\log n$, be as in Lemma 4.*

*With probability at least $1 - o(1)$ we have that for all $c \in [\sqrt{\log n}]$ there exists only one non-encoding $j := j_{\ell+c} \in [n] \setminus \mathcal{B}(x, y)$ with $y'_j = 1 - y^i_j$ for all $i \in [8e\log^{3/2}(n)/\log\log n]$ with $\operatorname{Lo}_{z,\sigma}(y^i) = \ell + c - 1$. Clearly, $j = \sigma(\ell + c)$.*

We are now ready to prove Theorem 1. As mentioned above, the proof also shows that the statement remains correct if we consider the unbiased black-box model with arbitrary arity.

*of Theorem 1.* We need to show that there exists an algorithm which maximizes any (a priori unknown) function $\text{Lo}_{z,\sigma} \in \textsc{LeadingOnes}_n$ using, on average, $O(n \log(n)/\log \log n)$ queries.

For readability purposes, let us fix some function $f = \text{Lo}_{z,\sigma} \in \textsc{LeadingOnes}_n$ to be maximized by the algorithm.

First, let us give a rough idea of our algorithm, Algorithm 4. A detailed analysis can be found below.

The main idea is the following. We maximize $f$ block-wise, where each block has a length of $\sqrt{\log n}$ bits. Due to the influence of the permutation $\sigma$ on $f$, these bit positions are a priori unknown. Assume for the moment that we have an $\ell$-encoding pair $(x, y)$, where $\ell \in [0..n - \lceil\sqrt{\log n}\,\rceil]$. In the beginning we have $\ell = 0$ and $y = x \oplus (1, \ldots, 1)$, the bitwise complement of $x$. To find an $(\ell + \sqrt{\log n})$-encoding pair, we first create a string $y'$ with objective value $f(y') \geq \ell + \sqrt{\log n}$. By Corollary 3, this requires on average $O(\log n)$ queries. Next, we need to identify the $\sqrt{\log n}$ $f(y)$-critical bit positions $\sigma(\ell + 1), \ldots, \sigma(\ell + \sqrt{\log n})$. To this end, we sample enough bit strings such that we can unambiguously identify these bit positions. As we shall see, this requires on average $O(\log^{3/2}(n)/\log \log n)$ queries. After identifying the critical bits, we update $(x, y)$ to a $(\ell + \sqrt{\log n})$-encoding pair. Since we need to optimize $n/\sqrt{\log n}$ such blocks of size $\sqrt{\log n}$, the overall expected optimization time is $O(n \log(n)/\log \log n)$.

Let us now present a more detailed analysis. We start by querying two complementary bit strings $x, y$. By swapping $x$ with $y$ in case $f(y) \geq f(x)$, we ensure that $f(x) > f(y) = 0$. This gives us a 0-encoding pair.

Let an $\ell$-encoding pair $(x, y)$, for some fixed value $\ell \in [0..n - \lceil\sqrt{\log n}\,\rceil]$, be given. We show how from this we find an $(\ell + \sqrt{\log n})$-encoding pair in an expected number of $O(\log^{3/2}(n)/\log \log n)$ queries.

As mentioned above, we first find a bit string $y'$ with objective value $f(y') \geq \ell + \sqrt{\log n}$. We do this by running Algorithm 3, the $(x, y)$-encoded $(1 + 1)$ EA with mutation probability $1/\sqrt{\log n}$ until we obtain such a bit string $y'$. By Corollary 3 this takes, on average, $O(\log n)$ queries.

Next we want to identify the $\sqrt{\log n}$ $\ell$-critical bit positions $\sigma(\ell+1), \ldots, \sigma(\ell+\sqrt{\log n})$. To this end, we query in the $i$-th iteration of the second phase, a bit string $y^i$ which has been created from $y'$ by flipping each non-encoding bit $y'_j$, $j \in [n] \setminus \mathcal{B}(x, y)$ independently with probability $1/\sqrt{\log n}$. If $f(y^i) = \ell + c - 1$ for some $c \in [\sqrt{\log n}]$, we update $X_{\ell+c} \leftarrow X_{\ell+c} \cup \{y^i\}$, the set of all queries with objective value $\ell + c - 1$, and we compute $\mathcal{J}_{\ell+c} := \{j \in [n] \setminus \mathcal{B}(x, y) \mid \forall w \in X_{\ell+c} : w_j = 1 - y'_j\}$, the set of candidates for $\sigma(\ell + c)$. We do so until we find $|\mathcal{J}_{\ell+c}| = 1$ for all $c \in [\sqrt{\log n}]$. By Corollary 6 this takes, on average, at most $8e \log^{3/2}(n)/\log \log n$ queries.

Thus, all we need to do in the third step is to update $(x, y)$ to an $(\ell + \sqrt{\log n})$-encoding pair by exploiting the information gathered in the second phase. For any $c \in [\sqrt{\log n}]$ let us denote the element in $\mathcal{J}_{\ell+c}$ by $j_{\ell+c}$. We go through the positions $\sigma(\ell+1), \ldots, \sigma(\ell+\sqrt{\log n})$ one after the other and either we update $y \leftarrow y \oplus e^n_{j_{\ell+c}}$ (if $f(y) < f(x)$), and we update $x \leftarrow x \oplus e^n_{j_{\ell+c}}$ otherwise. It is easy to verify that after $\sqrt{\log n}$ such steps we have $f(x) \geq \ell + \sqrt{\log n}$ and $f(y) \geq \ell + \sqrt{\log n}$. It remains to swap $(x, y) \leftarrow (y, x)$ in case $f(y) > f(x)$ in order to obtain an $(\ell + \sqrt{\log n})$-encoding pair $(x, y)$.

This shows how, given a $\ell$-encoding pair $(x, y)$, we find an $(\ell + \sqrt{\log n})$-encoding pair in $O(\log n) + O(\log^{3/2}(n)/\log \log n) + O(\sqrt{\log n}) = O(\log^{3/2}(n)/\log \log n)$ queries.

By definition of Algorithm 4, all bit positions in $\mathcal{B}(x, y)$ remain untouched in all further iterations of the algorithm. Thus, in total, we need to optimize $\lfloor \frac{n}{\lceil\sqrt{\log n}\,\rceil} \rfloor$ blocks of size $\lceil \sqrt{\log n}\, \rceil$ until we have a $(\lfloor \frac{n}{\lceil\sqrt{\log n}\,\rceil} \rfloor \lceil \sqrt{\log n}\, \rceil)$-encoding pair $(x, y)$. For each

**Algorithm 4:** A $*$-ary unbiased black-box algorithm for maximizing $f \in \text{LEADINGONES}_n$.
Recall that we have defined $\mathcal{J}_{\ell+c} := \{j \in [n] \backslash \mathcal{B}(x,y) \mid \forall w \in X_{\ell+c} : w_j = 1 - y'_j\}$.

**1 Initialization:**
**2**   for $i = 1, \ldots, n$ do $X_i \leftarrow \emptyset$;
**3**   Sample $x \in \{0,1\}^n$ uniformly at random;
**4**   Query $f(x)$;
**5**   Set $y \leftarrow x \oplus (1, \ldots, 1)$;
**6**   Query $f(y)$;
**7**   if $f(y) \geq f(x)$ then $(x,y) \leftarrow (y,x)$;
**8 Optimization:**
**9**   while $|\mathcal{B}(x,y)| \leq \lfloor \frac{n}{\lceil \sqrt{\log n} \rceil} \rfloor \lceil \sqrt{\log n} \rceil$ do
**10**      $\ell \leftarrow |\mathcal{B}(x,y)|$;
**11**      Apply Algorithm 3 with input $(x,y)$ and mutation probability $1/\sqrt{\log n}$ until it outputs a bit string $y'$ with $f(y') \geq \ell + \sqrt{\log n}$;
**12**      Initialize $i \leftarrow 1$;
**13**      while $\exists c \in [\sqrt{\log n}] : |\mathcal{J}_{\ell+c}| > 1$ do
**14**         $y^i \leftarrow \texttt{random}(y', x, y, 1/\sqrt{\log n})$;
**15**         Query $f(y^i)$;
**16**         if $f(y^i) \in [\ell, \ldots, \ell + \sqrt{\log n} - 1]$ then
**17**            Update $X_{f(y^i)+1} \leftarrow X_{f(y^i)+1} \cup \{y^i\}$;
**18**            Update $\mathcal{J}_{f(y^i)}$;
**19**         $i \leftarrow i + 1$;
**20**      for $c = 1, \ldots, \sqrt{\log n}$ do $\texttt{update}(x, y, y', X_{\ell+c})$;
**21**      if $f(y) > f(x)$ then $(x,y) \leftarrow (y,x)$;
**22**   Apply Algorithm 3 with input $(x,y)$ and mutation probability $1/\sqrt{\log n}$ until it queries for the first time a string $y'$ with $f(y') = n$;

block, the expected number of queries needed to fix the corresponding bit positions is $O(\log^{3/2}(n)/\log \log n)$. By linearity of expectation this yields a total expected optimization time of $O(n/\sqrt{\log n})O(\log^{3/2}(n)/\log \log n) = O(n \log(n)/\log \log n)$ for optimizing the first $k := \lfloor \frac{n}{\lceil \sqrt{\log n} \rceil} \rfloor \lceil \sqrt{\log n} \rceil$ bit positions $\sigma(1), \ldots, \sigma(k)$.

The remaining $n - k \leq \lfloor \sqrt{\log n} \rfloor$ bit positions can be found by Algorithm 3 in an expected number of $O(\log n)$ queries (Corollary 3). This does not change the asymptotic number of queries needed to identify $z$.

Putting everything together, we have shown that Algorithm 4 optimizes any function $\text{LO}_{z,\sigma} \in \text{LEADINGONES}_n$ in an expected number of $O(n \log(n)/\log \log n)$ queries. It is not difficult to verify that all variation operators are unbiased. We omit the details. □

## 4 The Unbiased Black-Box Complexity of LeadingOnes$_n$

Next we show how a slight modification of Algorithm 4 yields a 3-ary unbiased black-box algorithm with the same asymptotic expected optimization time.

**Theorem 7.** *The 3-ary unbiased black-box complexity of* $\text{LEADINGONES}_n$ *is* $O(n \log(n)/\log \log n)$.

**Algorithm 5:** Subroutine $\mathtt{update}(x, y, y', X_{\ell+c})$

**1 Input:** An $\ell$-encoding pair $(x, y)$, a bit string $y'$ with $f(y') \geq \ell + \sqrt{\log n}$, and, a set $X_{\ell+c}$ of samples $w$ with $f(w) = \ell + c - 1$ such that
$|\mathcal{J}_{\ell+c}| = |\{j \in [n] \backslash \mathcal{B}(x,y) \mid \forall w \in X_c : w_j = 1 - y'_j\}| = 1$;

**2 if** $f(y) \leq f(x)$ **then**
**3**   $\quad y \leftarrow y \oplus e^n_{\mathcal{J}(\ell+c)}$;
**4**   $\quad$ Query $f(y)$;
**5 else**
**6**   $\quad x \leftarrow x \oplus e^n_{\mathcal{J}(\ell+c)}$;
**7**   $\quad$ Query $f(x)$;

*Proof.* Key for this result is the fact that, instead of storing for any $c \in [\sqrt{\log n}]$ the whole query history $X_{\ell+c}$, we need to store only one additional bit string $x^{\ell+c}$ to keep all the information needed to determine $\sigma(\ell + c)$.

Algorithm 6 gives the full algorithm. Here, the bit string $\mathtt{update2}(w, y', x^{\ell+c})$ is defined via $\left(\mathtt{update2}(w, y', x^{\ell+c})\right)_i = w_i$ if $i \in [n] \backslash \mathcal{B}(y', x^{\ell+c})$ and $\left(\mathtt{update2}(w, y', x^{\ell+c})\right)_i = 1 - w_i$ for $i \in \mathcal{B}(y', x^{\ell+c})$.

Note that, throughout the run of the algorithm, the pair $(y', x^{\ell+c})$, or more precisely, the set $\mathcal{B}(y', x^{\ell+c})$ encodes which bit positions $j$ are still possible to equal $\sigma(\ell + c)$. Expressing the latter in the notation used in the proof of Theorem 1, we have in any iteration of the first **while**-loop that for all $i \in [n]$ it holds $y'_i = x^{\ell+c}_i$ if and only if $i \in \mathcal{J}_{\ell+c}$. This can be seen as follows. In the beginning, we only know that $\sigma(\ell+c) \neq \mathcal{B}(x,y)$. Thus, we initialize $x^{\ell+c}_i \leftarrow 1 - y'_i$ if $i \in \mathcal{B}(x,y)$ and $x^{\ell+c}_i \leftarrow y'_i$ for $i \in [n] \backslash \mathcal{B}(x,y)$. In each iteration of the second **while**-loop, we update $x^{\ell+c}_i \leftarrow 1 - y'_i$ if $\sigma(\ell + c) = i$ can no longer hold, i.e., if we have sampled a bit string $w$ with $f(w) = \ell + c - 1$ and $w_i = y'_i$.

It is easily verified that Algorithm 6 certifies Theorem 7. We omit a full proof in this extended abstract. □

## 5  LeadingOnes$_n$ in the Ranking-Based Models

As discussed above, we introduced two ranking-based versions of the black-box complexity notion in [DW11]: the *unbiased ranking-based* and the *unrestricted ranking-based black-box complexity*. Instead of querying the absolute fitness values $f(x)$, in the ranking-based model, the algorithms may only query the ranking of $y$ among all previously queried search points, cf. [DW11] for motivation and formal definitions. We briefly remark the following.

**Theorem 8.** *The 3-ary unbiased ranking-based black-box complexity of* LeadingOnes$_n$ *is* $O(n \log(n)/\log \log n)$.

This theorem immediately implies that the unrestricted ranking-based black-box complexity of LeadingOnes$_n$ is $O(n \log(n)/\log \log n)$ as well.

Theorem 8 can be proven by combining the Algorithm 6 presented in the proof of Theorem 7 with a sampling strategy as used in Lemma 4. Although the latter is not optimal, it suffices to show that after sampling $O(\log^{3/2}(n)/\log \log n)$ such samples, we can identify the rankings of $f(\ell+1), \ldots, f(\ell+\sqrt{\log n})$, with probability at least $1 - o(1)$. We do the sampling right after Line 10 of Algorithm 6. After having identified the rankings of $f(\ell+1), \ldots, f(\ell+\sqrt{\log n})$, we can continue as in Algorithm 6.

**Algorithm 6:** A 3-ary unbiased black-box algorithm for maximizing $f \in \text{LeadingOnes}_n$.

**1 Initialization:**
**2**   Sample $x \in \{0,1\}^n$ uniformly at random;
**3**   Query $f(x)$;
**4**   Set $y \leftarrow x \oplus (1, \ldots, 1)$;
**5**   Query $f(y)$;
**6**   if $f(y) \geq f(x)$ then $(x, y) \leftarrow (y, x)$;
**7 Optimization:**
**8**   while $|\mathcal{B}(x,y)| \leq \lfloor \frac{n}{\lceil \sqrt{\log n}\rceil} \rfloor \lceil \sqrt{\log n} \rceil$ do
**9**     $\ell \leftarrow |\mathcal{B}(x,y)|$;
**10**    Apply Algorithm 3 with input $(x, y)$ and mutation probability $1/\sqrt{\log n}$ until it outputs a bit string $y'$ with $f(y') \geq \ell + \sqrt{\log n}$;
**11**    for $c = 1, \ldots, \sqrt{\log n}$ do
**12**      for $i = 1, \ldots, n$ do
**13**        if $i \in \mathcal{B}(x,y)$ then $x_i^{\ell+c} \leftarrow 1 - y'_i$ else $x_i^{\ell+c} \leftarrow y'_i$;
**14**    while $\exists c \in [\sqrt{\log n}] : |\mathcal{B}(x^{\ell+c}, y')| > 1$ do
**15**      $w \leftarrow \texttt{random}(y', x, y, 1/\sqrt{\log n})$;
**16**      Query $f(w)$;
**17**      if $\exists c \in [\sqrt{\log n}] : f(w) = \ell + c - 1$ then
**18**        for $i = 1, \ldots, n$ do if $x_i^{\ell+c} = y'_i = w_i$ then $x_i^{\ell+c} \leftarrow 1 - y'_i$;
**19**    for $c = 1, \ldots, \sqrt{\log n}$ do
**20**      if $f(y) \leq f(x)$ then $\texttt{update2}(y, y', x^{\ell+c})$ else $\texttt{update2}(x, y', x^{\ell+c})$;
**21**    if $f(y) > f(x)$ then $(x, y) \leftarrow (y, x)$;
**22**  Apply Algorithm 3 with input $(x, y)$ and mutation probability $1/\sqrt{\log n}$ until it queries for the first time a string $y'$ with $f(y') = n$;

## 6 Conclusions

We have shown that there exists a 3-ary unbiased black-box algorithm which optimizes any function $\text{Lo}_{z,\sigma} \in \text{LeadingOnes}_n$ in an expected number of $O(n \log(n)/\log \log n)$ queries. This establishes a new upper bound on the unrestricted and the 3-ary unbiased black-box complexity of $\text{LeadingOnes}_n$.

Our result raises several questions for future research. The obvious one is to close the gap between the currently best lower bound of $\Omega(n)$ (cf. [DJW06]) and our upper bound of $O(n \log(n)/\log \log n)$. Currently, we cannot even prove an $\omega(n)$ lower bound. Secondly, it would also be interesting to know whether the gap between the 2-ary and the 3-ary unbiased black-box model is an artifact of our analysis or whether 3- and higher arity operators are truly more powerful than binary ones.

**Acknowledgments.** Carola Winzen is a recipient of the Google Europe Fellowship in Randomized Algorithms. This research is supported in part by this Google Fellowship.

# References

[AD11] Anne Auger and Benjamin Doerr. *Theory of Randomized Search Heuristics*. World Scientific, 2011.

[DJK$^+$11] Benjamin Doerr, Daniel Johannsen, Timo Kötzing, Per Kristian Lehre, Markus Wagner, and Carola Winzen. Faster black-box algorithms through higher arity operators. In *Proc. of Foundations of Genetic Algorithms (FOGA'11)*, pages 163–172. ACM, 2011.

[DJW02] Stefan Droste, Thomas Jansen, and Ingo Wegener. On the analysis of the (1+1) evolutionary algorithm. *Theoretical Computer Science*, 276:51–81, 2002.

[DJW06] Stefan Droste, Thomas Jansen, and Ingo Wegener. Upper and lower bounds for randomized search heuristics in black-box optimization. *Theory of Computing Systems*, 39:525–544, 2006.

[DW11] Benjamin Doerr and Carola Winzen. Towards a Complexity Theory of Randomized Search Heuristics: Ranking-Based Black-Box Complexity. In *Proc. of Computer Science Symposium in Russia (CSR'11)*, pages 15–28. Springer, 2011.

[LW10] Per Kristian Lehre and Carsten Witt. Black-box search by unbiased variation. In *Proc. of Genetic and Evolutionary Computation Conference (GECCO'10)*, pages 1441–1448. ACM, 2010.

[Müh92] Heinz Mühlenbein. How genetic algorithms really work: Mutation and hillclimbing. In *Proc. of Parallel Problem Solving from Nature (PPSN II)*, pages 15–26. Elsevier, 1992.

[Rud97] Günter Rudolph. *Convergence Properties of Evolutionary Algorithms*. Kovac, 1997.